\documentstyle[12pt]{article}

\begin{document}
\begin{titlepage}
\begin{center}
{\Large\bf On combining high and low $Q^2$ information \\
\vskip 0.5cm
on the polarized parton densities}
\end{center}
\vskip 2cm
\begin{center}
{\bf Elliot Leader}\\
{\it Birkbeck College, University of London\\
Malet Street, London WC1E 7HX, England\\
E-mail: E.Leader@bbk.ac.uk}\\
\vskip 0.5cm
{\bf Dimiter B. Stamenov \\
{\it Institute for Nuclear Research and Nuclear Energy\\
Bulgarian Academy of Sciences\\
blvd. Tsarigradsko Chaussee 72, Sofia 1784, Bulgaria\\
E-mail:stamenov@inrne.bas.bg\\
and\\
The Abdus Salam International Centre for Theoretical Physics,\\
Trieste, Italy}}
\end{center}

\vskip 0.3cm
\begin{abstract}

We draw attention to some problems in the combined use of high-$Q^2$
deep inelastic scattering (DIS) data and low-$Q^2$ hyperon 
$\beta$-decay data in the determination of the polarized parton 
densities. We explain why factorization schemes like the JET or AB 
schemes are the simplest in which to study the implications of the 
DIS parton densities for the physics of the low-$Q^2$ region.\\

\end{abstract}

\end{titlepage}

\newpage

\setcounter{page}{1}

{\large\bf 1. Introduction}
\vskip 4mm

Our most precise knowledge of the internal partonic structure of
the nucleon has come from decades of experiments on unpolarized
Deep Inelastic Scattering (DIS) of leptons on nucleons. More
recently there has been a dramatic improvement in the quality of
the data on polarized DIS and consequently an impressive growth
in the precision of our knowledge of the polarized parton
densities in the nucleon. However, it will be a long time before
the polarized data, for the moment limited to neutral current
reactions, can match the unpolarized data in volume and accuracy.
As a consequence, almost all analyses of the polarized parton
densities supplement the DIS (large $Q^2$) data with
information stemming from low-$Q^2$ weak interaction reactions.
More specifically, it is conventional to use the values of
$G_A/G_V$ from neutron $\beta$-decay, and 3F-D from hyperon
$\beta$-decays to help to pin down the values of the first
moments of certain combinations of parton densities.

However, the standard way of doing this has been criticized
because it essentially assumes exact $\rm SU(3)_f$ flavor
symmetry for the hyperon $\beta$-decays, whereas the growing 
precision of the measurements of magnetic moments and $G_A/G_V$
ratios in hyperon semi-leptonic decays may be indicating a non-
negligible breakdown of the flavor symmetry. Thus several attempts 
have been made \cite{Franklin}-\cite{Vogelsang} to incorporate some 
symmetry breaking in the combined analysis of weak interaction data 
and polarized DIS data.

We wish to point out in this note that there are inconsistencies in 
some of the schemes and to draw attention to certain 
essential requirements in any attempts to include $\rm SU(3)_f$ 
breaking in such combined analyses. We present also some comments 
regarding the question of the implications of the DIS parton 
densities for the physics of the low-$Q^2$ region.\\

{\large\bf 2. Some consequences of scheme dependence}
\vskip 4mm

From the measured spin asymmetries $A_{\parallel}$ and $A_{\bot}$ in 
the inclusive DIS of leptons on nucleons one obtains information 
on the spin structure function $~g_1(x,Q^2)~$ of the nucleon. 
In the next to leading order (NLO) QCD
approximation the quark-parton decomposition of $~g_1(x,Q^2)~$ has 
the following form:
\begin{equation}
g_1(x,Q^2)={1\over 2}\sum _{q} ^{N_f}e_{q}^2\left
[(\Delta q +\Delta\bar{q})\otimes (1 + {\alpha_s(Q^2)\over 2\pi}\delta 
C_q) +{\alpha_s(Q^2)\over 2\pi}\Delta G\otimes \delta C_G \right],
\label{g1partons} 
\end{equation}
where $\Delta q(x,Q^2), \Delta\bar{q}(x,Q^2)$ and $\Delta G(x,Q^2)$ are 
quark, anti-quark and gluon polarized densities which evolve in $Q^2$ 
according to the spin-dependent NLO DGLAP equations \cite{DGLAP}. 
($Q^2$ denotes the squared four-momentum of the exchanged virtual photon 
and $Q^2 > 1~GeV^2$ for DIS region.)
In (\ref {g1partons}) 
$\delta C_{q,G}$ are the NLO terms in the spin-dependent Wilson coefficient
functions and the symbol $\otimes$ denotes the usual convolution in Bjorken 
$x$ space. $N_f$ is the number of flavors.\\

In order to link the information on the polarized parton densities obtained 
from the DIS data with the information from  hyperon 
semi-leptonic weak decays it is convenient to rewrite (\ref {g1partons})
in terms of SU(3) flavor nonsinglet $\Delta q_{3,8}(x,Q^2)$ and singlet 
$\Delta \Sigma(x,Q^2)$ combinations of the quark densities ($N_f=3$):
\begin{eqnarray}
\nonumber
g_1^{p(n)}(x,Q^2)&=&{1\over 9} 
[(\pm {3\over 4}\Delta q_3 + {1\over 4}\Delta q_8 + \Delta \Sigma)\otimes 
(1 + {\alpha_s(Q^2)\over 2\pi}\delta C_q) \\
&+&{\alpha_s(Q^2)\over 2\pi}\Delta G\otimes \delta C_G]~,
\label{g1pn}
\end{eqnarray}
where
\begin{equation}
\Delta q_3(x,Q^2) = (\Delta u +\Delta\bar{u})(x,Q^2) - 
(\Delta d + \Delta\bar{d})(x,Q^2)~,
\label{delq3}
\end{equation}
\begin{eqnarray} 
\nonumber
\Delta q_8(x,Q^2)&=&(\Delta u +\Delta\bar{u})(x,Q^2) + 
(\Delta d + \Delta\bar{d})(x,Q^2) \\
&-&2(\Delta s+\Delta\bar{s})(x,Q^2)~,
\label{delq8}
\end{eqnarray}
\begin{equation}
\Delta \Sigma(x,Q^2)=(\Delta u +\Delta\bar{u})(x,Q^2)+
(\Delta d + \Delta\bar{d})(x,Q^2)+(\Delta s+\Delta\bar{s})(x,Q^2)~. 
\label{delsig} 
\end{equation}

Then for $\Gamma^{p(n)}_1(Q^2)$, the first moments of the proton and 
neutron structure functions $g_1^{p(n)}$, one has
\begin{eqnarray}
\nonumber
\Gamma^{p(n)}_1(Q^2)&=&\int _{0}^{1} dxg_1^{p(n)}(x,Q^2) \\
&=&{1\over 9}[\pm {3\over 4}a_3 + {1\over 4}a_8+ a_0(Q^2)] 
(1 - {\alpha_s(Q^2)\over \pi})~,
\label{Gamma1} 
\end{eqnarray}
where $a_3$ and $a_8$ are the nonsinglet axial charges corresponding 
to the $\rm 3^{rd}$ and $\rm 8^{th}$ components of the axial vector 
Cabibbo currents expressed in terms of the first
moments of the quark densities (\ref {delq3}) and (\ref {delq8})
[$\Delta q(Q^2)\equiv \int _{0}^{1} dx\Delta q(x,Q^2)$]
\begin{equation}
a_3 = (\Delta u +\Delta\bar{u})(Q^2) - (\Delta d + \Delta\bar{d})(Q^2)~,
\label{a3}
\end{equation}
\begin{equation} 
a_8=(\Delta u +\Delta\bar{u})(Q^2) + (\Delta d + \Delta\bar{d})(Q^2) 
-2(\Delta s+\Delta\bar{s})(Q^2)~.
\label{a8}
\end{equation}

Note that while $\Delta q$ and $\Delta\bar{q}$ depend on $Q^2$, $a_3$ 
and $a_8$ are conserved ($Q^2$ independent) quantities.

In (\ref {Gamma1}) $a_0(Q^2)$ is the singlet axial charge, which depends 
on $Q^2$ because of the axial anomaly. It must be emphasized that the 
connection between $a_0(Q^2)$ and the factorization scheme dependent 
quantity $\Delta \Sigma$, the first moment of the singlet quark density 
(\ref {delsig}), is different for the various factorization schemes used 
for the QCD calculations of the structure function $g_1$.

So, in the $\rm \overline{MS}$ scheme
\begin{equation}
a_0(Q^2) = \Delta \Sigma(Q^2)_{\rm \overline{MS}}~,
\label{a0MSbar}
\end{equation}
whereas for the JET and AB schemes in which $\Delta \Sigma$ is $Q^2$ 
independent,
\begin{equation}
a_0(Q^2) = \Delta \Sigma_{\rm JET(AB)}-N_f{\alpha_s(Q^2)\over 2\pi}
\Delta G(Q^2)_{\rm JET(AB)}~.
\label{a0JETAB}
\end{equation}

As pointed out in \cite{Zijlstra} one can actually define a family of such 
schemes. Among them the most popular are the so-called AB (Adler-Bardeen) 
\cite{ABscheme} and JET (see \cite{JETscheme} and references therein)
schemes. In (\ref{a0JETAB}) $\Delta G$ is the first moment of the 
polarized gluon density.
 
For the further considerations it is useful to recall the transformation 
rule connecting the first moments of the strange sea quarks in the nucleon,
$(\Delta s+\Delta\bar{s})$,
in the $\rm \overline{MS}$ and JET(AB) schemes:
\begin{equation}
(\Delta s+\Delta\bar{s})_{\rm JET(AB)}=
(\Delta s+\Delta\bar{s})(Q^2)_{\rm \overline{MS}} +
{\alpha_s(Q^2)\over 2\pi}\Delta G(Q^2)_{\rm \overline{MS}}~.
\label{delstrrule}
\end{equation}
Note that the LHS of (\ref {delstrrule}) is $Q^2$ independent.
 
It is important to mention here that the difference between the values of
$\Delta \Sigma$ or of $(\Delta s+\Delta\bar{s})$, obtained in the
$\rm \overline{MS}$ and JET(AB) schemes could be large due to the axial 
anomaly. Indeed, as a consequence of the anomaly, the term 
$\alpha_s\Delta G$ in (\ref{a0JETAB}) and (\ref{delstrrule})
behaves as \cite{EfrARCaCoMu}:  
\begin{equation}
\alpha_s(Q^2)\Delta G(Q^2)= \rm const +{\cal O}(\alpha_s(Q^2))~,
\label{aldelG}
\end{equation}
i.e., it is not really of order $\alpha_s$. 
To illustrate how large the difference can be, we present the values of
$(\Delta s+\Delta\bar{s})$ at $Q^2=1~GeV^2$ obtained in our recent 
analysis \cite{newanal} of the world DIS data in the $\rm \overline{MS}$ 
and JET(AB) schemes:
\begin{equation}
(\Delta s+\Delta\bar{s})_{\rm \overline{MS}}= -0.10 \pm 0.01,~~~~~
(\Delta s+\Delta\bar{s})_{\rm JET(AB)}= -0.06 \pm 0.01~.
\label{delsexp}
\end{equation}\\

\newpage
{\large\bf 3. What can be deduced in principle from DIS} 
\vskip 4mm

As was shown in \cite{spin98} if the DIS data on the independent structure 
functions $g_1^p$ and $g_1^n(g_1^d)$ were perfect and QCD was the correct 
theory of the strong interactions, the individual parton densities 
\begin{equation}
(\Delta u +\Delta\bar{u})(x,Q^2),~~~~
(\Delta d + \Delta\bar{d})(x,Q^2),~~~~ 
(\Delta s+\Delta\bar{s})(x,Q^2)
\label{indpart}
\end{equation}
and $~\Delta G(x,Q^2)~$ or, equivalently, $\Delta q_{3,8}(x,Q^2), 
\Delta \Sigma(x,Q^2)$ and $\Delta G(x,Q^2)$ would be {\it uniquely} 
determined at some arbitrary $Q^2=Q^2_0$. 
This follows from the fact that $\Delta q_3$ is fixed by the difference
$g_1^p-g_1^n$ while the rest $\Delta q_8, \Delta \Sigma$ 
and $\Delta G$ can be determined separately from $g_1^p+g_1^n$ because
of their {\it different} $Q^2$ evolution. 

It is immediately clear, given the limited range of $Q^2$
available and the fact that the data are {\it not} perfect and
have errors, that the separation of $\Delta q_{8},~\Delta \Sigma~$ and 
$~\Delta G$ from each other will not be very
clear-cut. Nonetheless, {\it in principle}, the data fix
$\Delta q_{3,8}(x,Q^2),~\Delta \Sigma(x,Q^2)~$ and 
$~\Delta G(x,Q^2)$ or, equivalently, via Eqs. (\ref{delq3}), (\ref{delq8}),
and (\ref{delsig}), $(\Delta u +\Delta\bar{u})(x,Q^2),
~(\Delta d + \Delta\bar{d})(x,Q^2),~(\Delta s+\Delta\bar{s})(x,Q^2)$ 
and $\Delta G(x,Q^2)$. 

It is also clear from the above that whereas the strange sea density
$(\Delta s+\Delta\bar{s})$ is, in principle, {\it fixed} by the inclusive 
(electromagnetic current) data, these data give no information about the 
other sea quark densities in the nucleon, $\Delta\bar{u}$ and 
$\Delta\bar{d}$, and therefore, about the valence parts
$~\Delta q_{v}~$ of the quark densities. In order to extract them from
the data (they are needed to make predictions for other processes, like 
polarized $pp$ reactions, etc.), additional assumptions about the 
flavor decomposition of the sea are necessary.
Conventionally, the following assumption has been used in most of 
the analyses 
\begin{equation}
\Delta\bar{u}=\Delta\bar{d}= \lambda \Delta\bar{s}~,
\label{SU3br}
\end{equation}
where $\lambda$ is a parameter. 

Given that the data fix $~\Delta q_{3,8},~\Delta \Sigma~$ and
$~\Delta G~$ and that
\begin{equation}
(\Delta s +\Delta\bar{s})(x,Q^2)={1\over 3}[\Delta \Sigma(x,Q^2) - 
\Delta q_8(x,Q^2)]~,
\label{s}
\end{equation}
we see that while $\Delta\bar{u}(\Delta u_v)$ and 
$\Delta\bar{d}(\Delta d_v)$
are sensitive to the assumptions about the flavor decomposition of the 
sea, the result for $(\Delta s +\Delta\bar{s})(x,Q^2)$ as well as for
$\Delta G(x,Q^2)$ should {\it not} change as $\lambda$ is varied. This 
provides a serious test for the stability of any analysis and was
confirmed numerically in our study \cite{spin98}.
(Note that the attempts \cite{semiincl} to extract the valence quarks from 
{\it semi-inclusive} data without assumptions about the sea are not 
entirely successful because of the quality of these data at present.)

In other words, inclusive DIS data do not enable us to test if the SU(3) 
symmetry 
of the sea is broken or not. What follows from these data and QCD is that  
$(\Delta s +\Delta\bar{s})(x,Q^2)$, the strange sea of the nucleon, does
{\it not} depend on the symmetry breaking of the sea and therefore, only
models, in which $~(\Delta s +\Delta\bar{s})(x,Q^2)~$ is {\it insensitive}
to this breaking, are acceptable.\\

{\large\bf 4. What we know about the partonic spin content of the nucleon 
from weak semi-leptonic hyperon decays}
\vskip 4mm

In addition to the information on the polarized parton densities from 
the DIS experiments very useful knowledge of their first moments comes 
from the hyperon semi-leptonic decays. 

The Bjorken sum rule \cite{BjoSR} tell us that
\begin{equation}
a_3 = (\Delta u +\Delta\bar{u})(Q^2) - (\Delta d + \Delta\bar{d})(Q^2)
={G_A\over G_V}(n\rightarrow p)\equiv g_A~,
\label{a3ga}
\end{equation}
where $g_A$ is the neutron weak $\beta$-decay constant \cite{PDG}:
\begin{equation}
g_A = 1.2601 \pm 0.0025~.
\label{ga}
\end{equation} 

This sum rule reflects the isospin SU(2) symmetry which is well established.
Assuming the usual SU(3) transformation properties of the axial currents and 
that the hyperons form an SU(3) octet, the 
hyperon $\beta$-decays fix $a_8$, the first moment of $\Delta q_8(x,Q^2)$,
to be:
\begin{equation} 
a_8=(\Delta u +\Delta\bar{u})(Q^2) + (\Delta d + \Delta\bar{d})(Q^2) 
-2(\Delta s+\Delta\bar{s})(Q^2)=\rm {3F-D}~,
\label{3FD}
\end{equation}
where
\begin{equation} 
\rm {3F-D} = 0.579 \pm 0.025~~\cite{ClRo}.
\label{3FDClRo}
\end{equation}
Depending on the data included in the hyperon $\beta$-decays analysis 
this value changes slightly: 
$0.601 \pm 0.038$ in \cite{Manohar} and $0.597 \pm 0.019$ in 
\cite{Ratcliffe}. 
However, the large value of $\chi^2/DOF$ of the SU(3) symmetric fit
(2.7 in \cite{Manohar} and 2.3 in \cite{Ratcliffe}) is some evidence for
SU(3) breaking. The issue of this breaking is treated in different models, 
which predict for $a_8$ values between 0.36 \cite{Manohar} and 0.85 
\cite{Gerasimov}.
The current KTeV experiment in Fermilab on the $\Xi^0~\beta$-decay,
$\Xi^0 \rightarrow \Sigma^{+}e\bar{\nu}$, (see \cite{RatST} and references
therein) will be a crucial test for them.

Given that because of the quality of the present DIS data the relations 
(\ref{a3ga}) and (\ref{3FD}) have been used in addition in most of
the analyses, it is quite important to understand what is the proper 
value of $a_8$. 

In this connection let us consider the  SU(3) symmetry breaking 
model suggested in Ref. \cite {3FDvalence}. In this model a specific
breaking of the symmetry in the hyperon decays is introduced by
treating the wave functions of the octet of baryons as made up of
a valence quark part, which is $\rm {SU(3)}_{f}$ symmetric, and a
sea part piece, which is allowed to break the symmetry. So, in this model 
Eqs. (\ref{a3ga}) and (\ref{3FD}) are modified in the following way
\begin{equation}
\Delta u_v (Q^2) - \Delta d_v(Q^2)=g_A,
\label{uvmindv}
\end{equation}
\begin{equation}
\Delta u_v (Q^2) + \Delta d_v(Q^2)=\rm {3F-D},
\label{uvplusdv}
\end{equation}
whereas for the sea quarks the assumption (\ref{SU3br}) has been used.
Note that while the sum rules (\ref{a3ga}) and (\ref{3FD}) are valid for 
any $Q^2$, the postulated relations (\ref{uvmindv}), (\ref{uvplusdv}) and 
(\ref{SU3br}) can only be valid, strictly speaking, at one value of $Q^2$, 
since the RH and LH sides evolve differently with $Q^2$. In the models of 
hyperon $\beta$-decays such relations must be imposed at very small 
$Q^2,~Q^2 \sim 0$, and thereafter cannot be assumed to hold in the DIS 
region. In particular, because of the anomalous gluon contribution, the 
equality (\ref{SU3br}) could be badly broken in the DIS region, especially 
in the $\rm \overline{MS}$-type scheme which is used in \cite{KL}. 
The misuse of (\ref{SU3br}) and (\ref{uvplusdv}) leads to {\it incorrect} 
expression for $a_8$ in the DIS region:
\begin{equation}
a_8 = \rm {3F-D} + 2\epsilon (\Delta s+\Delta\bar{s})(Q^2)~,
\label{a8KL}
\end{equation}
where $\epsilon =\lambda -1$ measures the breaking of the symmetry.

The use of Eq. (\ref{a8KL}) for $a_8$ in a combined analysis of the DIS 
and hyperon decays data is probably the origin of what we regard as a 
discrepancy in the results of \cite{KL}, namely that the strange quark 
content of the proton $(\Delta s+\Delta\bar{s})(Q^2)$ is sensitive to SU(3) 
breaking of the sea, in contradiction with the fact stressed in Section 3, 
that, at least in principle, the DIS data alone fix the value of the strange 
sea density $(\Delta s+\Delta\bar{s})(x,Q^2)$ and therefore, the strange sea 
quark polarization should not change as $\epsilon~(\lambda)$  is varied. \\

{\large\bf 5. How to link low $Q^2$ data with polarized DIS experiments}
\vskip 4mm

It is clear from the above that the polarized densities cannot be extracted 
well enough at present without linking the information from both high 
and low-$Q^2$ regions. In future more accurate both inclusive and 
semi-inclusive DIS data would help us to extract $a_8$ independently and 
thus to test models of the SU(3) flavor symmetry breaking. It also
appears that it might be possible in the not too distant future
to do high intensity neutrino experiments with a polarized target.\\

For the present we suggest the following regarding the combined use of 
DIS and low-$Q^2$ data and the question of the implications of the DIS 
parton densities for the physics of the low-$Q^2$ region. 

~~{\it i}) It is simplest to deal with quantities independent of $Q^2$.  
While $a_3$ and $a_8$ are independent of $Q^2$, the singlet combination of 
quarks $\Delta \Sigma(Q^2)$, as well as $(\Delta s+\Delta\bar{s})(Q^2)$, are 
in general $Q^2$ dependent. However, there are factorization schemes 
like the AB and JET schemes, in which $\Delta \Sigma(Q^2)$ and 
$(\Delta s+\Delta\bar{s})(Q^2)$ do {\it not} depend on $Q^2$. 
Although the theoretical results for the physical quantities such as 
the polarized structure functions $g_1^{p(n)}$ are scheme independent, 
it is clear that only in schemes like AB and JET, it is meaningful to 
directly interpret $\Delta \Sigma$ as the contribution of the quark spins to 
the nucleon spin and to confront its value obtained from the high energy 
region with predictions from low energy models like constituent quark, 
chiral quarks models, etc.

~{\it ii}) An important result from the present DIS data is that the singlet
axial charge is small:$a_0(Q^2)\approx 0.2-0.3$ in the high energy region 
$Q^2_{\rm DIS}\geq 1~GeV^2$. As a consequence, 
$(\Delta s+\Delta\bar{s})(Q^2)_{\rm \overline{MS}}$ 
in the $\rm \overline{MS}$ scheme is relatively large and negative in this 
region. This does not mean that 
$(\Delta s+\Delta\bar{s})(Q^2)_{\rm \overline{MS}}$
is large in the nonperturbative, low-$Q^2$ region too. 
To determine $(\Delta s+\Delta\bar{s})(Q^2)$ in the low-$Q^2$ region using 
the information on $a_0(Q^2)$ obtained in the DIS range, it is simplest 
to work in the JET(AB) scheme, in which
this quantity is $Q^2$ independent. As seen from Eqs. (\ref{delstrrule})
and (\ref{delsexp}), $(\Delta s+\Delta\bar{s})$ could be 
small, even zero (as expected in constituent quark models), depending on 
the sign and the
size of the gluon polarization in the DIS region. Although $\Delta G(Q^2)$
is not well determined from the present data there is significant evidence
from the experimental data,
directly \cite{HERMES} and indirectly via $Q^2$ evolution
effects (see, for example \cite{newanal}, \cite{gluon}), that 
$\Delta G(Q^2)$ is positive and not too small:
$\Delta G(Q^2)\sim 0.5-1.0$ at $Q^2 \sim 1~GeV^2$.

Here we would like to recall the interesting possibility of obtaining 
independent information on the strange sea quarks from the {\it elastic} 
$\nu(\bar{\nu})N$ scattering. As shown in the papers \cite{nuN}, 
measurements on the $\nu(\bar{\nu})$ asymmetry of these reactions will 
allow to extract $(\Delta s+\Delta\bar{s})(Q^2)$ at $Q^2\approx 0$ 
in a model-independent way.

{\it iii}) SU(3) symmetry breaking models invoked to explain hyperon decay 
data in the low-$Q^2$ range have to be consistent with the polarized 
parton densities and the QCD factorization scheme used in their 
determination from the DIS data.\\

{\large\bf 6. Summary}
\vskip 4mm

We have presented a critical assessment of what can be learned at 
present about the partonic spin content of the nucleon from both low 
energy hyperon $\beta$-decays and polarized DIS data. It was pointed 
out that the simplest way to link consistently the theoretical and 
experimental results 
obtained in these different regions is to use for the calculation of 
the spin-dependent structure function $g_1$ factorization schemes like
the JET and AB schemes. In these schemes, in addition to $a_3$ and $a_8$, 
the scheme dependent quantities $\Delta \Sigma$, 
$(\Delta s+\Delta\bar{s})$, the first 
moments of the singlet and strange sea densities, are also $Q^2$ {\it 
independent}. In the JET and AB schemes it is meaningful to interpret 
$\Delta \Sigma$ as the contribution of the quark spins to the nucleon 
spin and to compare its value obtained from DIS region with the 
predictions of the different (constituent, chiral, etc.) quark models at 
low $Q^2$. Finally, in such schemes the role of the gluon polarization in 
the high energy polarized experiments is more transparent. \\

{\large\bf Acknowledgments}
\vskip 4mm

One of us (D. S.) thanks S. B. Gerasimov for useful discussions 
concerning the hyperon $\beta$-decays. D. S. is grateful for the hospitality
of the High Energy Section of the Abdus Salam International Centre for
Theoretical Physics, Trieste, where this work has been completed. This 
research was partly supported by a UK Royal Society Collaborative Grant
and by the Bulgarian National Science Foundation.\\

\end{document}